\documentclass[twocolumn,aps,prl,superscriptaddress,showpacs,floatfix,longbibliography]{revtex4-1}
\usepackage{url}
\usepackage{cancel}
\usepackage[colorlinks,linkcolor=blue,citecolor=blue,filecolor=black,urlcolor=blue]{hyperref}
\usepackage{epsfig,graphics}
\usepackage{graphicx}% Include figure files
\usepackage{dcolumn}% Align table columns on decimal point
\usepackage{bm}% bold math
\usepackage[usenames]{color}
\usepackage{amssymb}
\usepackage{amsmath}
\usepackage{multirow}
\usepackage{float}
\usepackage{harpoon}
\usepackage{MnSymbol}
\usepackage{appendix}
\usepackage{color}
\usepackage{hyperref}
\usepackage{cleveref}

\newcommand{\sqrtsnn}{\mbox{$\sqrt{s_{\mathrm{NN}}}$}}
\newcommand{\pT} {p_{\mathrm{T}}}

\newcommand{\npart}{N_{\mathrm{part}}}

\usepackage{CJK}
\hfuzz=\maxdimen
\begin{document}
\title{Evidence of quadrupole and octupole deformations in $^{96}$Zr+$^{96}$Zr and $^{96}$Ru+$^{96}$Ru collisions at ultra-relativistic energies}
\newcommand{\sbu}{Department of Chemistry, Stony Brook University, Stony Brook, NY 11794, USA}
\newcommand{\bnl}{Physics Department, Brookhaven National Laboratory, Upton, NY 11976, USA}
\author{Chunjian Zhang}\affiliation{\sbu}
\author{Jiangyong Jia}\email[Correspond to\ ]{jiangyong.jia@stonybrook.edu}\affiliation{\sbu}\affiliation{\bnl}

\begin{abstract}
In the hydrodynamic model description of heavy ion collisions, the elliptic flow $v_2$ and triangular flow $v_3$ are sensitive to the quadrupole deformation $\beta_2$ and octupole deformation $\beta_3$ of the colliding nuclei. The relations between $v_n$ and $\beta_n$ have recently been clarified and were found to follow a simple parametric form. The STAR Collaboration have just published precision $v_n$ data from isobaric $^{96}$Ru+$^{96}$Ru and $^{96}$Zr+$^{96}$Zr collisions, where they observe large differences in central collisions $v_{2,\mathrm{Ru}}>v_{2,\mathrm{Zr}}$ and $v_{3,\mathrm{Ru}}<v_{3,\mathrm{Zr}}$. Using a transport model simulation, we show that these orderings are a natural consequence of $\beta_{2,\mathrm{Ru}}\gg\beta_{2,\mathrm{Zr}}$ and $\beta_{3,\mathrm{Ru}}\ll\beta_{3,\mathrm{Zr}}$. We reproduce the centrality dependence of the $v_2$ ratio qualitatively and $v_3$ ratio quantitatively, and extract values of $\beta_2$ and $\beta_3$ that are consistent with those measured at low energy nuclear structure experiments. STAR data provide the first direct evidence of strong octupole correlations in the ground state of $^{96}$Zr in heavy ion collisions. Our analysis demonstrates that flow measurements in high-energy heavy ion collisions, especially using isobaric systems, are a new precision tool to study nuclear structure physics.
\end{abstract}
%	25.75.Gz, Particle correlations and fluctuations
%	25.75.Ld,	Collective flow, relativistic collisions.
%      25.75.-q,	Relativistic heavy-ion collisions                
\maketitle

Most atomic nuclei present intrinsic deformed shapes, characterized notably by quadrupole, octupole and hexadecapole components~\cite{Heyde2011,Heyde:2016sop}. Experimental evidences for nuclear deformation are primarily extracted from spectroscopic measurements and models of reduced transition probability between low-lying rotational states, which involves nuclear experiments with energy per nucleon less than few 10 MeVs. Recently, the prospects of probing the nuclear deformation at much higher beam energy, energy per nucleon exceeding hundreds of GeVs, by taking advantage of the responses of hydrodynamic collective flow of the final state particles to the shape and sizes of the initial state, have been discussed~\cite{Heinz:2004ir,Filip:2009zz,Shou:2014eya,Goldschmidt:2015kpa,Giacalone:2017dud,Giacalone:2018apa,Giacalone:2021uhj,Giacalone:2021udy,Jia:2021wbq,Jia:2021tzt,Bally:2021qys}, and several experimental evidences have been observed~\cite{Adamczyk:2015obl,Acharya:2018ihu,Sirunyan:2019wqp,Aad:2019xmh,jjia}. 

Nuclear deformation is often described by a nucleon density in a deformed Woods-Saxon form:
\begin{align}\nonumber
\rho(r,\theta,\phi)&=\frac{\rho_0}{1+e^{[r-R(\theta,\phi)]/a}},\\\label{eq:f1}
 R(\theta,\phi) &=R_0\left(1+\beta_2 Y_2^0 +\beta_3 Y_3^0+\beta_4 Y_4^0\right),
\end{align}
where the nuclear surface $R(\theta,\phi)$ includes only the most relevant axial symmetric quadrupole, octupole and hexadecapole deformations~\footnote{Including other internal shape degrees of freedom, such as triaxiality, does not affect the two-particle observables $\varepsilon_n^2$ and $v_n^2$~\cite{Jia:2021tzt}.}.

It is straightforward to see why dynamics of heavy-ion collisions is sensitive to nuclear deformation. These high-energy collisions deposit a large amount of energy in the overlap region, forming a hot and dense quark-gluon plasma (QGP)~\cite{Busza:2018rrf}, whose initial shape in the transverse plane is sensitive to the nuclear deformation. This initial shape is characterized via eccentricities $\varepsilon_n= |\int r^n e^{in\phi} e(r,\phi) rdrd\phi/\int r^n e(r,\phi) rdrd\phi|$, estimated from the energy density $e(r,\phi)$ in the overlap. Driven by the pressure gradient forces and subsequent hydrodynamic collective expansion, the initial $\varepsilon_n$ are transferred into azimuthal anisotropy of final state hadrons~\cite{Heinz:2013wva},  dominated by an elliptic and an triangular modulation of particle distribution, $dN/d\phi\!\propto\!1+\!2v_2 \cos2(\!\phi-\Psi_2\!)+\!2v_3\cos3(\!\phi-\Psi_3\!)$. The $v_n$ coefficients reflect hydrodynamic response of the QGP to $\varepsilon_n$, and follow an approximate linear relation $v_n=k_n\varepsilon_n$ for events in fixed centrality~\cite{Niemi:2015qia,Schenke:2020uqq}. In collisions of spherical nuclei, $\varepsilon_2$ mainly reflects the elliptic shape of the overlap controlled by the impact parameter, while non-zero $\varepsilon_3$ arises from random fluctuations of nucleons. The presence of non-zero $\beta_n$ enhances $\varepsilon_n$, and consequently the values of $v_n$, which on general ground follow a simple quadratic form~\cite{Jia:2021qyu} for $n=2$ and 3.
\begin{align}\label{eq:f2}
\varepsilon_n^2\!=\!a_n'\!+\hspace{-1.5mm}\sum_{m,k=2}^{4}\hspace{-1.5mm}b_{n;mk}'\beta_{m}\beta_{k},\;\;\;v_n^2\!=\!a_n+\hspace{-1.5mm}\sum_{m,k=2}^{4}\hspace{-1.5mm}b_{n;mk}\beta_{m}\beta_{k}, 
\end{align}
where $\varepsilon_n^2$ and $v_n^2$ are mean-square values calculated for events in a narrow centrality. The $a_n'$ and $a_n$ are values for collisions of spherical nuclei, which are strong functions of system size and centrality. In contrast, the coefficients $b'$ and $b$ are expected to be nearly independent of system size~\cite{Jia:2021tzt,Jia:2021qyu}. This is because deformations change the distribution of nucleons in the entire nucleus, the coefficients in a give centrality depend mainly on the alignment two nuclei and centrality, not the system size.

All the $b'$ and $b$ terms are not equally important. In fact in the ultra-central collisions (UCC), model studies show that only one deformation term is important~\cite{Jia:2021tzt},
\begin{align}\label{eq:f3}
\varepsilon_n^2 =a_n'+b_{n}'\beta_{n}^2\;, \hspace{20pt} v_n^2=a_n+b_{n}\beta_{n}^2.
\end{align}
These simple relations provide a powerful data-driven method to constrain the $\beta_2$ and $\beta_3$ by comparing collisions of two species with similar sizes, see Refs.~\cite{Giacalone:2021udy,Jia:2021wbq,Jia:2021tzt} for details. The most straightforward scenario is to consider collisions of two isobaric systems $X+X$ and $Y+Y$ with the same mass number, therefore having the same coefficients in Eq.~\ref{eq:f3}. In this case, these ratios have a particularly simple expression,
\begin{align}\nonumber
\frac{\varepsilon_{n,Y}^2}{\varepsilon_{n,X}^2} &= 1+\frac{b_n'}{a_n'}(\beta_{n,Y}^2-\beta_{n,X}^2)/(1+\frac{b_n'}{a_n'}\beta_{n,X}^2)\\\label{eq:f4}
                                              &\approx  1+\frac{b_n'}{a_n'}(\beta_{n,Y}^2-\beta_{n,X}^2)\;,\\\nonumber
\frac{v_{n,Y}^2}{v_{n,X}^2} &= 1+\frac{b_n}{a_n}(\beta_{n,Y}^2-\beta_{n,X}^2)/(1+\frac{b_n}{a_n}\beta_{n,X}^2)\\\label{eq:f5}
                          &\approx 1+\frac{b_n}{a_n}(\beta_{n,Y}^2-\beta_{n,X}^2)\;.
\end{align}
The relative ordering of $v_n$ in the two systems is a direct reflection of the ordering of their $\beta_n$ values~\cite{Giacalone:2021udy}. Besides, in the UCC region, the values of $a_n'$ and $a_n$ are smallest and the influence of deformations is largest.

In this Letter, we apply this idea to the recent isobar $^{96}_{40}$Zr+$^{96}_{40}$Zr and $^{96}_{44}$Ru+$^{96}_{44}$Ru collision~\cite{STAR:2021mii}, and make predictions on ratios of $v_n$ from the known values of $\beta_2$ and $\beta_3$ from nuclear structure measurements, as given in Table~\ref{tab:1}. Assuming the same $\beta_n$ for neutrons and protons and a uniform charge distribution out to the distance $R(\theta,\phi)$, the $\beta_n$ values are obtained from the measured reduced electric transition probability $B(En)\!\uparrow$ via the standard formula~\cite{bohr}, 
\begin{align}\label{eq:f6}
\beta_2=\frac{4\pi}{3 Z R_{0}^{2}}\sqrt{\frac{B(E 2)\!\uparrow}{\mathrm{e}^{2}}}\;,\;\beta_3=\frac{4\pi}{3 Z R_{0}^{3}}\sqrt{\frac{B(E 3)\!\uparrow}{\mathrm{e}^{2}}},
\end{align}
with $R_0=1.2A^{1/3}$ fm. We note that the number of neutrons in $^{96}$Zr nucleus is equal to one of the so-called ``octupole magic'' numbers 56~\cite{Butler:1996zz}. Low energy experiments indeed show that $^{96}$Zr has a very large octupole collectivity corresponding to a large $\beta_{3}$ value, but a small $\beta_2$ value~\cite{Stautberg:1967hrb,Mach:1990zz,Hofer:1993gou}. On the other hand, the $^{96}$Ru nucleus has larger $\beta_2$, but shows no evidence of significant $\beta_3$. The latter is expected from the very large excitation energy for its $3_1^-$ state. In this analysis, we assume $\beta_{3,\mathrm{Ru}}=0$. Note that the predictions from nuclear structure models~\cite{Moller:2015fba} have large discrepancy from these data, therefore they are not used in this study. 

\begin{table}[!h]
\centering
\begin{tabular}{|c|cc|cc|}\hline
          &$\beta_2$  &$E_{2_1^+}$ (MeV) &$\beta_3$  & $E_{3_1^-}$ (MeV)\\[1ex]\hline
$^{96}$Ru &$0.154$     & 0.83            & -   &  3.08\\[1ex]\hline
$^{96}$Zr &$0.062$     & 1.75            &0.202,0.235,0.27 &  1.90\\[1ex]\hline
\end{tabular}
\caption{\label{tab:1} Values of $\beta_2$ deduced from the $B(E2;0_1^+\rightarrow 2_1^+)$~\cite{Pritychenko:2013gwa} and $\beta_3$ deduced from three $B(E3;0_1^+\rightarrow 3_1^-)$~\cite{Kibedi:2002ube} transition measurements via Eq.~\ref{eq:f6}. The corresponding excitation energies are also provided. In general, larger excitation energy corresponds to smaller deformability for the nucleus.} 
\end{table}

To understand the hydrodynamic response to nuclear deformations and make predictions,  we employ the multi-phase transport model (AMPT)~\cite{Lin:2004en} as a proxy for hydrodynamics. This model is successful in describing collective flow data in small and large collision systems at RHIC and LHC~\cite{Adare:2015cpn,Ma:2014pva,Bzdak:2014dia,Nie:2018xog} and has been used to study the $\beta_n$ dependence of $v_n$~\cite{Giacalone:2021udy,Jia:2021wbq}. We use AMPT v2.26t5 in string-melting mode at $\sqrtsnn=200$~GeV, and a partonic cross section of 3.0~$m$b~\cite{Ma:2014pva,Bzdak:2014dia}, which gives a reasonable description of Au+Au and U+U $v_2$ data at RHIC~\cite{Xu:2011fe,STAR:2021twy}.  We simulate generic isobar $^{96}$X+$^{96}$X collisions with $R_0=5.09$ fm and $a=0.52$ fm. We performed a scan on $\beta_2$: $\beta_2=0,0.05,0.1,0.15,0.2$ and $\beta_3=0$, as well as a scan on $\beta_3$: $\beta_3=0,0.05,0.1,0.15,0.2$ and $\beta_2=0.06$. The $\varepsilon_n$ are calculated from participating nucleons and $v_n$ are calculated using two-particle correlation method with hadrons in $0.2<\pT<2$ GeV and $|\eta|<2$~\cite{ATLAS:2012at}.

\begin{figure}[!h]
\includegraphics[width=0.985\linewidth]{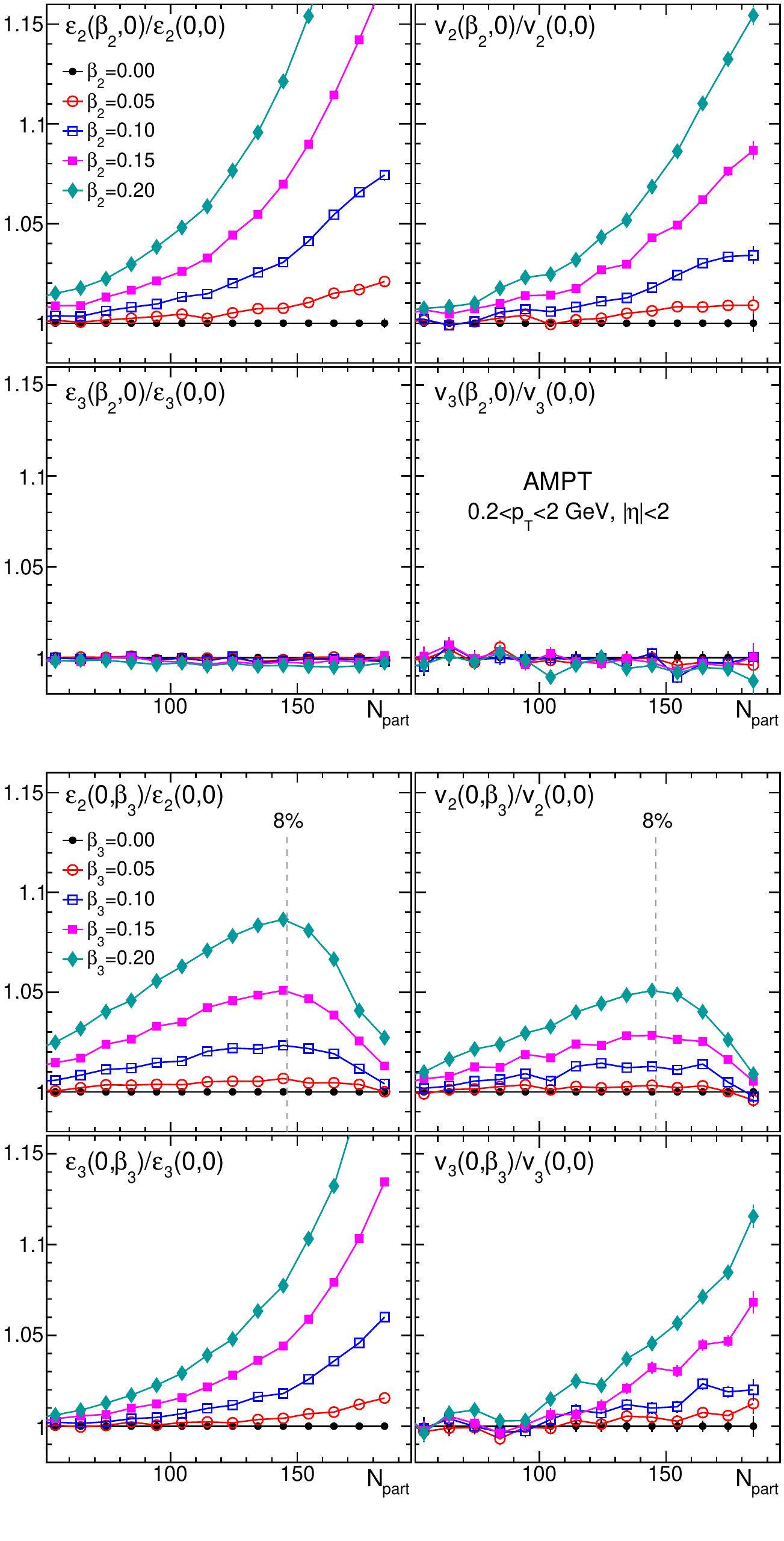}
\caption{\label{fig:1} The $\npart$ dependence of ratios of $\varepsilon_n$ (left column) or $v_n$ (right column) relative to undeformed case with for $n=2$,3 in generic $^{96}$X+$^{96}$X collisions with different $\beta_2$ (top part) and $\beta_3$ (bottom part) values. The $v_n$ are calculated in the AMPT models with hadrons in $|\eta|<2$ and $0.2 < \pT < 2$ GeV.}
\end{figure}

The left panels of Fig.~\ref{fig:1} show ratios of $\varepsilon_n(\beta_2,\beta_3)$ for given values of $\beta_2$ or $\beta_3$ to that for spherical nuclei. There are four different types of ratio considered, $\varepsilon_3(\beta_2,0)/\varepsilon_2(0,0)$, $\varepsilon_3(0,\beta_3)/\varepsilon_3(0,0)$,  $\varepsilon_2(0,\beta_3)/\varepsilon_2(0,0)$, and $\varepsilon_3(\beta_2,0)/\varepsilon_3(0,0)$, i.e. we not only consider how the $\varepsilon_n$ are affected by $\beta_n$ but also the cross-correlation between $\varepsilon_2$ and $\beta_3$ or between $\varepsilon_3$ and $\beta_2$. Our study shows that the $\varepsilon_n$ in Glauber model takes the following simplified version of Eq.~\ref{eq:f2}~\cite{Jia:2021tzt},
\begin{align}\label{eq:f7}
\varepsilon_2^2=a_2'+b_2'\beta_2^2+b_{2,3}'\beta_3^2,\;\;
\varepsilon_3^2=a_3'+b_3'\beta_3^2.
\end{align}
The $\varepsilon_2$ is strongly influenced by $\beta_3$ in the non-central collisions, reaching a maximum at $\npart\sim146$ corresponding to about 8\% centrality. The right panels of Fig.~\ref{fig:1} show ratios of $v_n^2$ with the same layout, which we found can be well parametrized by
\begin{align}\label{eq:f8}
v_2^2=a_2+b_2\beta_2^2+b_{2,3}\beta_3^2,\;\;
v_3^2=a_3+b_3\beta_3^2.
\end{align}

It is clear from Fig.~\ref{fig:1} that the ratio of $v_n$ in all cases is smaller than the ratio of $\varepsilon_n$ for the same $\beta_n$. This observation implies that the hydrodynamic response for $b'$ parameters are smaller than those for the $a'$, i.e. $b_n/b_n'<a_n/a_n'$ and $b_{2,3}/b_{2,3}'<a_2/a_2'$. In other words, the contributions of $v_n$ from nuclear deformation are more damped by hydrodynamic evolution than that from the eccentricity of the spherical nuclei. One possible explanation is that the nuclear deformation increases $\varepsilon_n$ by adding more nucleons towards the edge of the overlap, leading to a weaker hydrodynamic response and less efficient conversion of $\varepsilon_n$ to $v_n$. 

With Eq.~\ref{eq:f7} in hand and using the values of $\beta_n$ from Tab.~\ref{tab:1}, we are ready to predict the $v_n$ ratios between Ru+Ru and Zr+Zr collisions. Safely assuming $\beta_{3,\mathrm{Ru}}=0$, these ratios are expected to scale like:
\begin{align}\label{eq:f9}
\frac{v_{2,\mathrm{Ru}}^2}{v_{2,\mathrm{Zr}}^2}&\approx 1+\frac{b_2}{a_2}(\beta_{2,\mathrm{Ru}}^2-\beta_{2,\mathrm{Zr}}^2)-\frac{b_{2,3}}{a_2}\beta_{3,\mathrm{Zr}}^2,\\\label{eq:f10}
\frac{v_{3,\mathrm{Ru}}^2}{v_{3,\mathrm{Zr}}^2}&\approx 1-\frac{b_3}{a_3}\beta_{3,\mathrm{Zr}}^2.
\end{align}
The coefficients $b_n/a_n$ and $b_{2,3}/a_2$ can be calculated from Fig.~\ref{fig:1} as a function of $\npart$, which can then be used to predict $v_{n,\mathrm{Ru}}/v_{n,\mathrm{Zr}}$. Alternatively, $v_{n,\mathrm{Ru}}/v_{n,\mathrm{Zr}}$ can be obtained directly according to the $\beta_n$ values in Tab.~\ref{tab:1},
\small{\begin{align}\nonumber
\frac{v_{n,\mathrm{Ru}}}{v_{n,\mathrm{Zr}}}\! &\!\approx\!\frac{v_{n}(\!\beta_{2,\mathrm{Ru}},0\!)}{v_{n}(\!\beta_{2,\mathrm{Zr}},0\!)}\!\times\!\frac{v_{n}(\!0,\beta_{3,\mathrm{Ru}}\!)}{v_{n}(\!0,\beta_{3,\mathrm{Zr}}\!)}\!=\!\frac{v_{n}(\!0.154,0\!)}{v_{n}(\!0.062,0\!)}\!\times\!\frac{v_{n}\!(0,0)}{v_{n}\!(0,0.2\!)}\;.
\end{align}\normalsize
%\frac{v_{3,\mathrm{Ru}}}{v_{3,\mathrm{Zr}}}\!&\!\approx\! \frac{v_{3}(0,\beta_{3,\mathrm{Ru}}\!)}{v_{3}(0\!,\beta_{3,\mathrm{Zr}}\!)}\!\times\!\frac{v_{3}(\beta_{2,\mathrm{Ru}},0\!)}{v_{3}(\beta_{2,\!Zr},0)}\!=\!\frac{v_{3}(0,0)}{v_{3}(0,0.2\!)}\!\times\!\frac{\!v_{3}(0.154,0\!)}{\!v_{3}(0.062,0\!)}.

Since $\beta_{2,\mathrm{Ru}}\gg\beta_{2,\mathrm{Zr}}$ and value of $\beta_{3,\mathrm{Zr}}$ is large, the ratio $v_{2,\mathrm{Ru}}/v_{2,\mathrm{Zr}}$ is expected to contain a positive contribution from $\beta_2$ and a negative contribution from $\beta_{3}$. This is clearly demonstrated in the top panel of Fig.~\ref{fig:2}. The $\beta_{3,\mathrm{Zr}}$ has a small impact in the 0--1\% centrality, but significantly reduces the $v_2$ ratio over the 1--40\% centrality range, with the maximum reduction at around $\npart\sim 146$ or 8\% centrality. The influence of $\beta_{3,\mathrm{Zr}}$ also forces a much sharper decrease of the $v_2$ ratio in the centrality range of 0--5\%, and leads to a non-monotonic centrality dependence. On the other hand, the predicted $v_{3,\mathrm{Ru}}/v_{3,\mathrm{Zr}}$ ratios in the bottom panel of Fig.~\ref{fig:2} are independent of the $\beta_{2}$ of the two systems. This interesting interplay between $\beta_2$ and $\beta_3$, and the resulting features in the $v_n$ ratio in the isobar collisions are salient and robust predictions that can be verified experimentally.

The STAR Collaboration has just released the $v_{n,\mathrm{Ru}}/v_{n,\mathrm{Zr}}$ data in several coarse centrality bins~\cite{STAR:2021mii}; they are contrasted with our predictions in Fig.~\ref{fig:2}. The $v_2$ ratio data show non-monotonic centrality dependence similar in shape to our prediction that include effects of both $\beta_2$ and $\beta_3$, and such non-monotonicity was not predicted in previous studies that did not include the influence of $\beta_3$~\cite{Deng:2016knn,Xu:2021vpn}. However our prediction is systematically lower than the STAR data by up to 2\% in the mid-central and peripheral region. This residual difference could be due to the large neutron skin effect in $^{96}$Zr~\cite{Xu:2021vpn}, which was found to enhance $\varepsilon_{2,\mathrm{Ru}}/\varepsilon_{2,\mathrm{Zr}}$ with a shape and magnitude similar to this difference. Remarkably, our prediction of $v_3$ ratio agrees nearly perfectly with the STAR data over the entire centrality range when $\beta_{3,\mathrm{Zr}}=0.2$ is used. This also raise a question whether the large $\beta_3$ value constrained by the isobar collisions could imply a large static octupole deformation in the ground state of $^{96}$Zr, as implemented in our AMPT model simulation.

\begin{figure}[!h]
\includegraphics[width=0.8\linewidth]{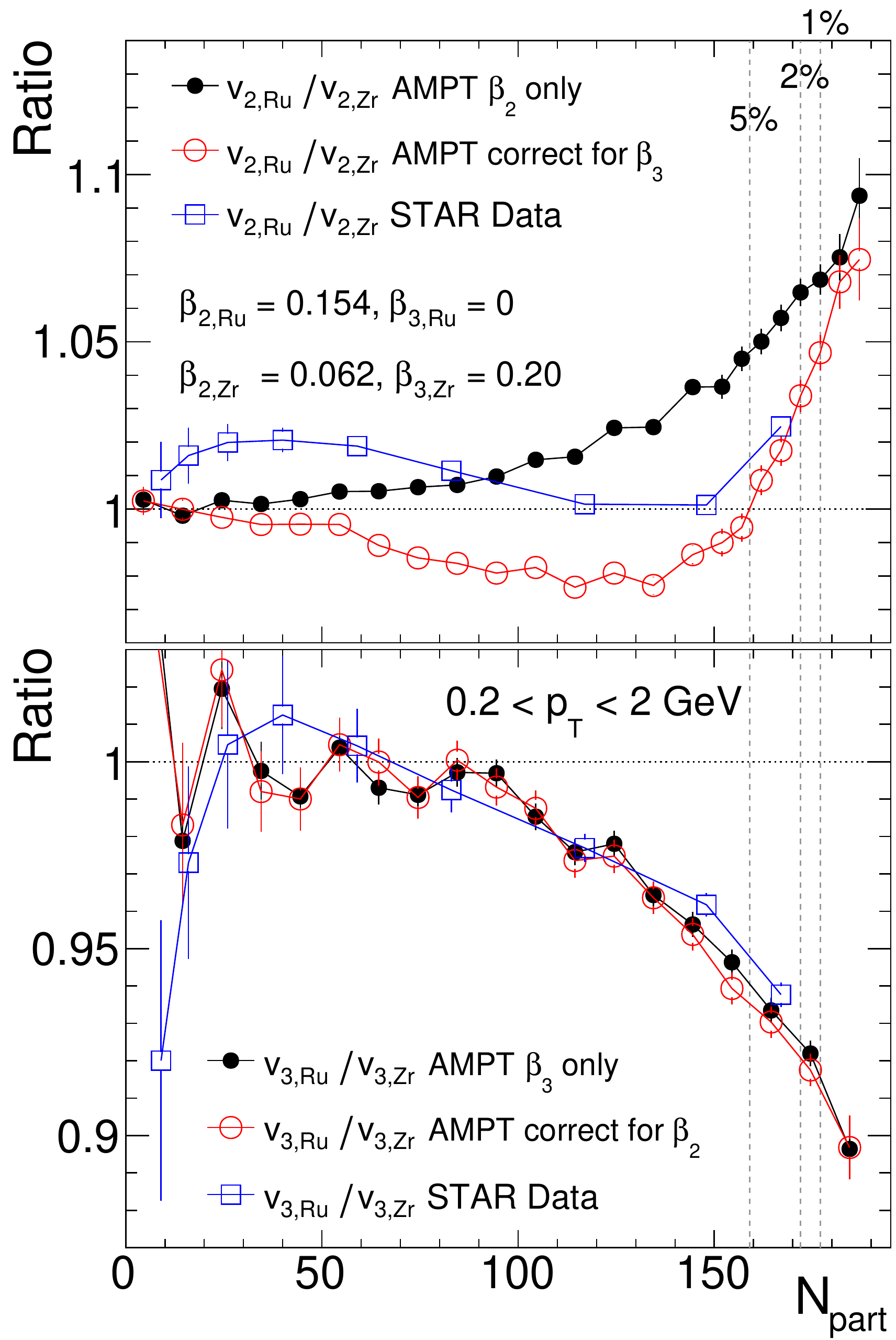}
\caption{\label{fig:2} Predicted ratios of $v_2$ (top) and $v_3$ (top) between Ru+Ru and Zr+Zr collisions from the AMPT model based on the $\beta_2$ and $\beta_3$ values for Ru and Zr from Table~\ref{tab:1}. They are compared with the STAR data from Ref.~\cite{STAR:2021mii}.}
\end{figure}

Next, we make predictions in the 0--1\% most central collisions, where the influence of nuclear deformation is strongest, the contribution from $\beta_3$ to $v_2$ is small, and $v_n$ ratios are expected to be described precisely by Eq.~\ref{eq:f5} with $b_n/a_n$ as the only free parameters. The STAR data are not yet available in this region. Figure~\ref{fig:3} shows the predicted ratio $v^2_{2,\mathrm{Ru}}/v^2_{2,\mathrm{Zr}}$ and $v^2_{3,\mathrm{Ru}}/v^2_{3,\mathrm{Zr}}$ as a function of $\beta_{2,\mathrm{Ru}}^2$ and $\beta_{3,\mathrm{Zr}}^2$, respectively. The predictions follow strictly a linear dependence on $\beta_n^2$. Assuming the $\beta_{2,\mathrm{Ru}}=0.154$ and $\beta_{2,\mathrm{Zr}}=0.062$ from Table~\ref{tab:1}, we predicts a $v^2_{2,\mathrm{Ru}}/v^2_{2,\mathrm{Zr}}\approx 1.16$, corresponding to $v_{2,\mathrm{Ru}}/v_{2,\mathrm{Zr}}\approx 1.08$. On the other hand, the three octupole deformation values from nuclear structure measurements, $\beta_{3,\mathrm{Zr}}= 0.202, 0.235, 0.27$, would predict $v^2_{3,\mathrm{Zr}}/v^2_{3,\mathrm{Ru}}=1.24,1.33,1.44$ as indicated by the solid arrows in Fig.~\ref{fig:3}, or equivalently $v_{3,\mathrm{Ru}}/v_{3,\mathrm{Zr}}=0.90,0.87,0.83$. The latter two cases, $\beta_{3,\mathrm{Zr}}=0.235$ and 0.27, are clearly ruled out by the data-theory comparison of the $v_3$ ratio in the bottom panel of Fig.~\ref{fig:2}. These two $\beta_{3,\mathrm{Zr}}$ values would also lead to further reduction of the $v_{2,\mathrm{Ru}}/v_{2,\mathrm{Zr}}$ ratio (indicated by the open circles) in the top panel of Fig.~\ref{fig:2}. This additional reduction amount to $0.235^2/0.2^2-1=38\%$ or $0.27^2/0.2^2-1=82\%$ of the difference between the two AMPT predictions, which will make both the shape and the magnitude of the $v_{2,\mathrm{Ru}}/v_{2,\mathrm{Zr}}$ incompatible with the experimental data even after including the predicted neutron skin effects~\cite{Xu:2021vpn}. Therefore, for the first time, the isobar collisions show clear potential for providing new constraints on nuclear deformation parameters that can complement those from nuclear structure spectroscopy in testing state-of-the-art nuclear structure models. In order to pin down the interplay between $\beta_2$ and $\beta_3$ more quantitatively, however, STAR measurement~\cite{STAR:2021mii} should be extended to finer centrality bins, especially in the 0--5\% range where the $v_n$ ratios are still changing very rapidly.
% in resolving discrepancies of $\beta_3$ from low energy experiments. 

\begin{figure}[!h]
\includegraphics[width=0.985\linewidth]{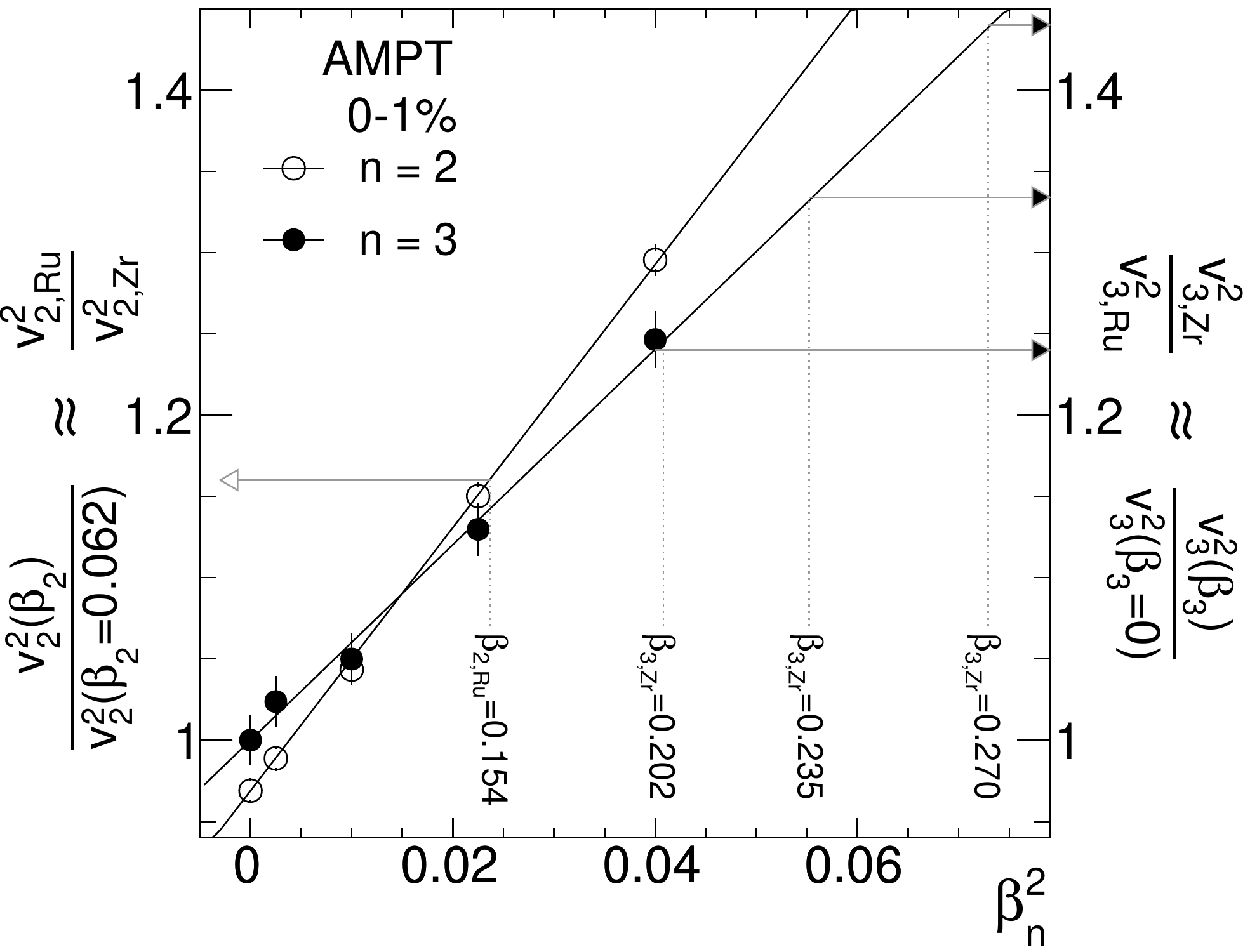}
\caption{\label{fig:3} Predicted ratio of $v_{2,\mathrm{Ru}}^2/v_{2,\mathrm{Zr}}^2$ as a function of $\beta_{2,\mathrm{Ru}}^2$ assuming $\beta_{2,\mathrm{Zr}}=0.062$ (open symbols), and ratio of $v_{3,\mathrm{Zr}}^2/v_{3,\mathrm{Ru}}^2$ as a function of $\beta_{3,\mathrm{Zr}}^2$ assuming $\beta_{3,\mathrm{Ru}}=0$ (filled symbols) from the AMPT model for the 0--1\% centrality. The arrows indicate the corresponding ratios for $\beta_{2,\mathrm{Ru}}=0.154$ (left pointing) and $\beta_{3,\mathrm{Zr}}=0.202,0.235$ and 0.27 from Table~\ref{tab:1} (right pointing).}
\end{figure}

A few homeworks are required in order to accomplish a precision determination of nuclear deformations using heavy ion collisions. First and most importantly, we need to understand the connection of the $\beta_n$ given by nuclear structure method Eq.~\ref{eq:f6} and $\beta_n$ in heavy ion collisions via Eq.~\ref{eq:f1}. The former measures the charge distribution at the timescale of $10^{-21}$s, while the heavy ion collisions cares only about the mass distribution at a much shorter time of $10^{-24}$s, which is also strongly Lorentz contracted unlike in the low-energy experiments.  For example, changing the $R_0$ value alone in Eq.~\ref{eq:f6} would directly change the value of extracted $\beta_n$, but changing the $R_0$ in Eq.~\ref{eq:f1} has little impact on the hydrodynamic response to $\beta_n$. From the modeling side of the heavy ion collisions, we need to pin down the hydrodynamic response of $v_n$ to the deformation contribution to the eccentricity, which is clearly different from the hydrodynamic response of $v_n$ to the baseline eccentricity for spherical nuclei (see Fig.~\ref{fig:1}). Further hydrodynamic models studies are required to quantify the systematic uncertainties, in particular the ratio $a_n/b_n$. Nevertheless, the STAR results and our study demonstrate that heavy ion collisions can serve as a new tool for imaging the shape of the atomic nuclei and possibly other features of their nuclear structure. This provides good arguments for an extended system scan of different isobaric systems for precision measurement of interesting nuclear structure effects and complement the low energy methods~\cite{Jia:2021wbq,Jia:2021tzt}. 

In summary, recent STAR measurement of $v_2$ and $v_3$ show significant differences between $^{96}$Zr+$^{96}$Zr and $^{96}$Ru+$^{96}$Ru collisions. Using a transport simulation, we show that these differences can be naturally explained from the large quadrupole deformation $\beta_2$ of $^{96}$Ru and large octupole deformation $\beta_3$ of $^{96}$Zr. Our calculations quantitively describe the $v_{3,\mathrm{Ru}}/v_{3,\mathrm{Zr}}$ by assuming $\beta_{3,\mathrm{Zr}}=0.2$ from one nuclear structure measurement. However, the presence of $\beta_{3,\mathrm{Zr}}$ was found to significantly enhance the $v_{2,\mathrm{Zr}}$ and led to a non-monotonic centrality dependence of $v_{2,\mathrm{Ru}}/v_{2,\mathrm{Zr}}$ as observed in the data. Additional physics such as neutron skin differences are also required to quantitively describe the $v_2$ ratio. Our analysis demonstrates that isobaric heavy ion collisions can be used as a precision tool to image the shape and radial structures of the nuclei. We hope this can be done in the existing high-energy collider facilities in the near future.

{\bf Acknowledgements:} We thank Yu Hu for providing the STAR data. We thank Giuliano Giacalone for proof reading the draft and his valuable suggestions. We acknowledge Somadutta Bhatta for discussions during the development of this work. This work is supported by DOE DEFG0287ER40331.

\bibliography{deform}{}
\end{document}